\documentclass[aps,pra,twocolumn,groupedaddress,showpacs]{revtex4}

\usepackage{graphicx}
\usepackage{amsmath}
\usepackage{bm}
\usepackage{color}

\newcommand{\Eq}[1]{Eq.~(\ref{#1})}
\newcommand{\Fig}[1]{Fig.~\ref{#1}}
\newcommand{\Ref}[1]{Ref.~\cite{#1}}
\def\beq{\begin{equation}} \def\eeq{\end{equation}}
\def\bea{\begin{eqnarray}} \def\eea{\end{eqnarray}}
\def\bse{\begin{subequations}} \def\ese{\end{subequations}}

\def\||{\parallel}
\let\p\partial

\def\<{\left\langle} \def\>{\right\rangle}
\def\({\left(} \def\){\right)}
\def\[[{\left[} \def\]]{\right]}

\begin{document}


\title{Nambu--Goldstone Modes in Segregated Bose--Einstein Condensates}


\author{Hiromitsu Takeuchi$^{1}$}
\author{Kenichi Kasamatsu$^{2}$}
\affiliation{
$^1$Department of Physics, Osaka City University, Sumiyoshi-ku, Osaka 558-8585, Japan \\
$^2$Department of Physics, Kinki University, Higashi-Osaka, 577-8502, Japan }


\date{\today}

\begin{abstract}
Nambu--Goldstone modes in immiscible two-component Bose-Einstein 
condensates are studied theoretically.
 In a uniform system,
 a flat domain wall is stabilized and then the translational invariance normal to the wall is spontaneously broken in addition to the breaking of two U(1) symmetries in the presence of two complex order parameters. 
We clarify properties of the low-energy excitations and identify that there exist two Nambu--Goldstone modes: an in-phase phonon with a linear dispersion and a ripplon with a fractional dispersion.
The signature of the characteristic dispersion can be verified in segregated condensates in a harmonic potential.
\end{abstract}


\maketitle
Spontaneous symmetry breaking (SSB) is a universal phenomenon 
that can occur in almost all energy scales in nature. 
When continuum symmetries of a system are spontaneously broken, there exist gapless modes 
called Nambu--Goldstone (NG) modes \cite{Nambu}, which determine the low-energy 
properties of the system. 
In Lorenz-invariant systems, the number of NG modes ($n_{\rm NG}$)
 coincides with the number of broken symmetries ($n_{\rm SSB}$),
 and their dispersions are linear as $\epsilon \propto p$,
 where $\epsilon$ and ${\bm p}$ are the energy and momentum of the NG modes, respectively.
However, this counting rule of $n_{\rm NG}$ fails in systems without Lorentz invariance 
or with spontaneously broken spacetime symmetry \cite{Nielsen1976, Low2002, Nambu2004, Brauner}. 
Recently, the NG theorem was generalized to the nonrelativistic system 
and the counting rule has been 
re-examined \cite{Watanabe2011, Watanabe, Hidaka, Watanabe2}.

Superfluids or Bose--Einstein condensates (BECs) \cite{Donnelly, Pethickbook} are important foundations upon which to study SSB and NG modes.
In a single-component superfluid or a scalar BEC, the appearance of the order parameter is associated with the SSB 
of the global U(1) symmetry. The accompanying NG mode is known as a phonon with a linear dispersion.
Another important example is a spinor BEC with spin-1 bosons \cite{Kawaguchi},
where the symmetry of spin rotation is also broken in addition to the global U(1).
 In the polar phase with $n_{\rm NG}=n_{\rm SSB}=3$, the dispersions of the three NG modes are all linear.
However, in the ferromagnetic phase, we have only two NG modes with $n_{\rm SSB}=3$ and $n_{\rm NG}=2$,
 and then a NG mode has a linear dispersion and the other has a quadratic one
 that comes from a conjugate pairing between the generators of the two broken symmetries \cite{Watanabe}.

The NG theorem is nontrivial
 when the spatial symmetry of the system is spontaneously broken.
Such examples can be seen in a scalar BEC with vortices.
When there is a straight vortex along the $z$ axis in a uniform system,
the translational symmetries in the $x$ and $y$ directions are explicitly broken, and thus $n_{\rm SSB}=3$.
However, there are only two NG modes: 
the Kelvin mode with a quadratic dispersion, which causes helical deformation of a vortex line, 
 and the varicose mode with a linear dispersion, which corresponds to a phonon propagating along the vortex core \cite{Donnelly,Takeuchi2009, Kobayashi2013}.
Another example can be seen in Tkachenko modes of two-dimensional vortex lattices in rotating superfluids \cite{Watanabe2}.

A nontrivial example can also be seen in two-component BECs, i.e., condensates of two distinguishable bosons \cite{Kasarev}.  
The system is characterized by two-component order parameters $\Psi_j~(j=1,2)$, which 
allow the excitation of two independent phonons associated with the two broken U(1) symmetries.
Two-component BECs are characterized by the inter-atomic coupling constants 
$g_{jk}$ proportional to the $s$-wave scattering length $a_{jk}$ between the $j$th and $k$th components.
A homogeneous two-component system is miscible for $ g_{12}^2 < g_{11}g_{22}$ with $g_{jj}>0$ and then there are two NG modes with $\epsilon \propto p$, while for the ``fine-tuned" interaction parameter $g_{12}^2=g_{11}g_{22}$ the dispersion of a mode becomes quadratic with $\epsilon \propto p^2$. 
For $g_{12}>\sqrt{g_{11}g_{22}}$, the strong inter-component repulsion leads to a 
phase-separation forming domain walls \cite{Tim,Ao,Trippenbach,Schaeybroeck}.
In the presence of a flat domain wall, the translational symmetry normal to the wall is spontaneously broken,
and thus the transverse shift of the wall costs zero energy. 
There are some studies on interface modes in segregated two-component BECs \cite{Mazets,Sasaki2009,Takeuchi,Kobyakov2011,Roy}. 
Mazets showed that, under the Bogoliubov--de Gennes (BdG) analysis with a suitable 
ansatz of the domain wall profile, there are two branches of waves localized 
near a domain wall with fractional dispersions, i.e., 
$\epsilon \propto p^{3/2}$ and $\epsilon \propto p^{1/2}$ \cite{Mazets}. 
Later, the dispersion $\epsilon \propto p^{3/2}$ was derived from the hydrodynamic effective theory of ripple 
waves on a domain wall in segregated BECs \cite{Takeuchi}.
However, the number of NG modes and their details in the low-energy limit have not been confirmed yet.

In this paper, we provide a full account of the NG modes in segregated two-component BECs.
The semi-classical analysis of the NG modes in a uniform system reveals that 
the dispersion $\epsilon \propto p^\gamma$ with $\gamma <1$ is forbidden for the localized modes 
in the low-energy limit, which conflicts with the prediction of $\gamma=1/2$ in \Ref{Mazets}. 
Even in the highly compressible BECs, the interface mode, referred to as a ripplon, has 
a dispersion $\epsilon \propto p^{3/2}$, which is similar to a standard capillary 
wave in classical incompressible hydrodynamics. 
We confirm that, despite $n_{\rm SSB} = 3$, there exist two NG modes, a ripplon and a phonon, in the low-energy limit 
by carefully calculating the system size dependence of the dispersion relation through the numerical analysis of the BdG equation.
We also discuss these low-energy modes in segregated condensates trapped by a harmonic potential.

We consider binary BECs described by complex order parameters $\Psi_j = \sqrt{n_j} e^{i \theta_j}$ in the mean-field approximation at zero temperature. 
Let us start with the Gross-Pitaevskii Lagrangian for the two-component BECs, 
\begin{equation}
{\cal L} = \int d^3x \left( {\cal P}_1 + {\cal P}_2 - g_{12} |\Psi_1|^2 |\Psi_2|^2
\right), \label{lagra}
\end{equation}
where
$
{\cal P}_j =
 i \hbar \Psi_j^* \partial_t \Psi_j + (\hbar^2/2 m_j) \Psi_j^* \nabla^2 \Psi_j 
- (V_j-\mu_j) |\Psi_j|^2  -g_{jj} |\Psi_j|^4/2
$
with $m_j$, $V_j$, and $\mu_j$ being the atomic mass, the external trap potential, and 
the chemical potential of the $j$th component, respectively.
The intra- and inter-component interaction parameters have the form
$g_{jk} = 2\pi \hbar^2 a_{jk}(m_j^{-1} + m_k^{-1})$. We assume $a_{jk} > 0$ in the following. 

We first consider a homogeneous system with $V_j=0$.
The condensates are miscible and immiscible when $g_{11}g_{22} > g_{12}^2$ and $g_{11}g_{22} < g_{12}^2$, respectively \cite{Tim,Ao}.
For both cases, there occurs the trivial SSB related to the phases $\theta_1$ and $\theta_2$;
the Lagrangian is invariant when the phases are rotated independently.
For the immiscible case, the spatial symmetry is additionally broken in the presence 
of a domain wall due to the phase separation. 
This symmetry breaking may cause a ripplon, which is a quantum representation of ripple 
waves propagating on a domain wall, as a NG mode in this system. 
We confine ourselves to the immiscible case throughout this work.

\begin{figure}
\begin{center}
\includegraphics[width=.98 \linewidth,keepaspectratio]{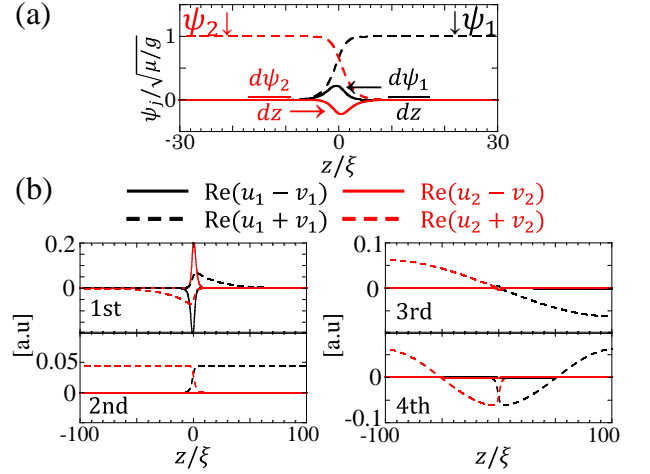}
\end{center}
\caption{(Color online) (a) Profiles of order parameters $\psi_j$ and their spatial derivatives in a stationary state of segregated condensates with a domain wall.
(b) Typical profiles of the NG modes (first: ripplon, second: in-phase phonon) and the massive modes (third, fourth) with $p\xi/\hbar=0.05$.
 The solid and dashed curves represent the density [${\rm Re}(u_j-v_j)$] and phase [${\rm Re}(u_j+v_j)$] fluctuations, respectively,  with ${\rm Im}u_j\approx 0$ and ${\rm Im}v_j\approx 0$.} 
\label{domainprofile}
\end{figure}

Assuming a flat domain wall normal to the $z$ axis in a stationary state $\Psi_j=\psi_j(z)$, 
Eq.~(\ref{lagra}) yields
\begin{equation}
0 = \( -\frac{\hbar^2}{2m_j} \frac{d^2}{dz^2} 
-\mu_j+ \sum_{k} g_{jk} \psi_k^2 \) \psi_j.  
\label{staredeqeq}
\end{equation}
Here, we have assumed ${\rm Im}\psi_j=0$ without loss of generality.
Figure \ref{domainprofile}(a) shows the spatial profile of the stationary solution 
with a flat domain wall for the feasible parameters 
$\mu_1 = \mu_2 = \mu$, $g_{11} = g_{22}=g$, $m_1 = m_2=m$ and $g_{12}/g=1.2$, e.g., in Ref. \cite{Tojo2010}.
The length is scaled by the healing length $\xi=\hbar/\sqrt{m\mu}$.
The two components are separated by the domain wall along the $z = 0$ plane.
There is an overlapping region, in which one component penetrates into
the other component.
The width $\xi_w$ of the overlapping region diverges as 
$\xi_w \propto (1-g_{12}/\sqrt{g_{11}g_{22}})^{-1/2}$ for $g_{12}\to\sqrt{g_{11}g_{22}}$ \cite{Ao}.
Here, we consider a well-segregated case with $\xi_w \sim \xi$.
The tension of a flat domain wall is calculated as the excess energy in the presence of a domain wall.
$ \sigma=-\sum_j (\hbar^2/4m_j)\int dz \psi_j\p_z^2\psi_j$, under the pressure equilibrium $P_1=P_2$ 
with the hydrostatic pressure $P_j=\mu_j^2/2g_{jj}=P$ of the $j$th component in the bulk, $|z| \to \infty$ \cite{Ao}.

The NG modes in the segregated BECs are fully solved by the BdG formalism. 
By writing a perturbation of the order parameters as
\bea
\delta\Psi_j=u_j^n(z)e^{i ({\bm p}\cdot {\bm x}-\epsilon_n t)/\hbar}-v_j^n(z)^*e^{-i ({\bm p}\cdot {\bm x}-\epsilon_n t)/\hbar}
\label{eq:fluctuation}
\eea
 with ${\bm p}=(p_x,p_y,0)^T$, one obtains the BdG equations, 
\bea
\epsilon_n {\bm u}_n=
 \hat{h}\hat{\rho}_3{\bm u}_n,
\label{eq:reducedBdG}
\eea
where we used ${\bm u}_n=(u_1^n, u_2^n, v_1^n, v_2^n)^T$, $\hat{\rho}_3={\rm diag}(1,1,-1,-1)$,
\bea
\hat{h}=\(
\begin{array}{cc}
{\cal H}+{\cal G} & {\cal G}\\
{\cal G} & {\cal H}+{\cal G} \\
\end{array} 
\)
\eea
with $[{\cal G}]_{jk}=g_{jk}\psi_j\psi_k$, and ${\cal H}={\rm diag}({\cal H}_1,{\cal H}_2)$ with
${\cal H}_j= {\bm p}^2/2m_j-\hbar^2 \partial_z^2 / 2m_j+\sum_k g_{kj}\psi_k^2-\mu_j$. 
The real quantities $\epsilon_n$ and ${\bm p}$ correspond to the quantum 
of energy and momentum of the mode when the Bogoliubov coefficients 
are normalized as ${\cal N}_{nn}=1$, where 
\bea
{\cal N}_{nn'} \equiv \int d^3x {\bm u}_n^\dagger \hat{\rho}_3{\bm u}_{n'}. 
\label{eq:normal}
\eea
The existence of the zero modes for $p=|{\bm p}|=0$ is easily confirmed 
because Eq.~(\ref{eq:reducedBdG}) is satisfied for the zero-energy perturbations 
\bea
u_j=-v_j^*=i \frac{\Delta \Theta_j}{2} \psi_j
\label{eq:zero_phase}
\eea
and
\bea
u_j=-v_j^*=\frac{\Delta z}{2} \frac{d\psi_j}{dz}
\label{eq:zero_shift}
\eea
 with small real constants $\Delta\Theta_j$ and $\Delta z$,
which are connected with the independent phase rotations [$\psi_j \to e^{i\Delta\Theta_j} \psi_j$] and the translational shift of the domain wall [$\psi_j(z) \to \psi_j(z+\Delta z)$], respectively.

It is useful to discuss some restriction for NG modes from their asymptotic behavior 
for $|z| \gg \xi_w$.
The asymptotic profiles of the Bogoliubov coefficients $u_j$ and $v_j$ are
obtained by a method similar to the semi-classical approximation in 
quantum mechanics \cite{Landau} 
by introducing an effective semi-classical action $S$ in the form
\bea
 {\bm u}(z)={\bm U} e^{iS(z)/\hbar}
 \label{eq:semi-U}
\eea
with ${\bm U}=[{\cal U}_1,{\cal U}_2,{\cal V}_1,{\cal V}_2]^T$.
In the semi-classical limit $\hbar\to 0$, which is well applicable 
in the bulk region with $d\psi_j/dz \to 0$, one obtains from Eq. (\ref{eq:reducedBdG})
 \bea
 \frac{dS}{dz}=\pm \sqrt{\frac{\epsilon^2}{c_j^2}-p^2}
 \label{eq:dS}
 \eea
with the sound velocity $c_j=\sqrt{\mu_j/m_j}$ ($j=1$ for $z\to -\infty$ and $j=2$ for $z\to \infty$) in the bulk \cite{Res2}.

Consider the NG modes $\epsilon \propto cp^\gamma$ for $p\to 0$ 
with a positive constant $c$ and an exponent $\gamma$.
In segregated BECs, the NG modes are classified into the bulk NG mode (BNG mode) and the localized NG mode (LNG mode).
The former has finite coefficients $u_j$ and/or $v_j$ in the bulk.
Since the amplitude $|\psi_1|$ ($|\psi_2|$) vanishes for $z\to \infty$ ($z\to -\infty$), a possible BNG mode is a phonon in a single-component BEC: 
 \bea
 \epsilon=c_jp~~~~({\rm BNG~mode}).
 \eea
The localized NG mode satisfies the condition that $u_j$ and $v_j$ vanish far from the domain wall 
$z\to \pm \infty$. Then, the exponent $S$ of Eq.~(\ref{eq:semi-U}) should be imaginary and one obtains 
in the low-momentum limit as
\bea
S=\left\{
\begin{array}{ll}
i(-1)^j \sqrt{1-(c/c_j)^2}pz & (\gamma=1) \\
i(-1)^j pz & (\gamma >1) \\
\end{array}
\right.~~({\rm LNG~mode}),
\eea
where we neglected the integration constants and used $j=1$ ($2$) for $z\to-\infty$ ($+\infty$) and $c<c_j$ for $\gamma=1$.
Thus, $\gamma=1/2$ found in Ref.~\cite{Mazets} is excluded for LNG modes.
The localized mode with $\gamma=1/2$, an analog of surface gravity wave in hydrodynamics, can appear in the presence of external potentials \cite{Res1}.

To find a possible value of $\gamma$ in the above semi-classical analysis,
we analyze the low-energy effective theory describing the LNG mode (ripplon) by taking into account the zero mode due to a transverse shift of the wall. 
Here, we do not remove the condition of compressibility in the BECs, while 
the previous works analyzed this problem based on the assumption 
of incompressibility \cite{Sasaki2009,Kobyakov2011}. 
Neglecting the internal structure of the domain wall and representing the position of the domain wall as $z=\eta(x,y,t)$,
we may approximate Eq.~(\ref{lagra}) to the effective Lagrangian
\bea
{\cal L}_{\rm eff}=\int dxdy \(    \int^\eta_{-\infty}dz {\cal P}_1+\int_\eta^{+\infty}dz {\cal P}_2- \sigma{\cal S}  \)
\eea
with ${\cal S}=\sqrt{1+(\p_x \eta)^2+(\p_y \eta)^2 }$ and the domain wall tension coefficient $\sigma$.
The variation of ${\cal L}_{\rm eff}$ about small $\eta$ gives
\bea
{\cal P}_1(\eta)-{\cal P}_2(\eta)+\sigma (\p_x^2+\p_y^2)\eta=0.
\label{eq:Bernoulli}
\eea
 The density and phase perturbations due to a small fluctuation Eq.(\ref{eq:fluctuation}) are written respectively as
$\delta n_j \approx \delta\bar{n}_j \cos({\bm p}\cdot{\bm x}/\hbar - \epsilon t/\hbar + \alpha_j)$
 and
$\delta \theta_j \approx \delta\bar{\theta}_j \sin({\bm p}\cdot{\bm x}/\hbar - \epsilon t/\hbar + \beta_j)$
with $\delta\bar{n}^2=4\[[ ({\rm Re} u_j-{\rm Re} v_j)^2+({\rm Im} u_j+{\rm Im} v_j)^2 \]]|\psi_j|^2$, $\delta\bar{\theta}_j^2=\[[ ({\rm Re} u_j+{\rm Re} v_j)^2+({\rm Im} u_j-{\rm Im} v_j)^2 \]]/|\psi_j|^2$, $\cos\alpha_j=2{\rm Re}(u_j-v_j)|\psi_j|/\delta\bar{n}_j$, and $\sin \beta_j={\rm Im}(u_j-v_j)/|\psi_j|\delta\bar{\theta}_j$.
For the LNG modes, one obtains with \Eq{eq:semi-U} for $|z| \gg \xi_w$ ,
\bea
&&\delta n_j = 2\psi_je^{a_j p z/\hbar}({\cal U}_j-{\cal V}_j)\cos({\bm p}\cdot {\bm x}/\hbar-\epsilon t/\hbar),
\label{eq:d_n}\\
&&\delta \theta_j = \psi_j^{-1} e^{a_jp z/\hbar}({\cal U}_j+{\cal V}_j)\sin({\bm p}\cdot {\bm x}/\hbar-\epsilon t/\hbar),
\label{eq:d_theta}
\eea
with ${\rm Im} {\cal U}_j={\rm Im} {\cal V}_j=0$, $a_j=-(-1)^j$ for $\gamma >1$, and $a_j=-(-1)^j\sqrt{1-c^2/c_j^2}$ for $\gamma=1$.

In the low-momentum limit $p\to 0$, $\delta n_j$ vanishes because ${\cal U}_j/{\cal V}_j=\[[ cp^\gamma+(1-a_j^2)p^2/2m_j+\mu_j\]]/\mu_j\to 1$.
This justifies the incompressibility-like assumption $\delta n_j=0$ in Refs. \cite{Sasaki2009, Kobyakov2011} exactly in the low-energy limit even for highly compressible gaseous BECs.
Additionally, we impose the kinematic boundary condition, $\frac{\hbar}{m_j}\p_z \delta\theta_j=\p_t \eta$, which means that a superfluid velocity 
in the $z$ direction on the wall causes a shift of the wall.
By combining this condition with \Eq{eq:Bernoulli}, we obtain the dispersion for a ripple wave $\eta\propto \cos({\bm p}\cdot {\bm x}/\hbar-\epsilon t/\hbar)$,
\bea
\epsilon=\epsilon_{\rm rip}(p)\equiv\sqrt{\frac{\sigma/\hbar}{\rho_1+\rho_2}} p^{3/2}
\label{eq:dis_ripplon}
\eea
with $\rho_j=m_j\mu_j/g_{jj}$.

\begin{figure}
\begin{center}
\includegraphics[width=0.98 \linewidth,keepaspectratio]{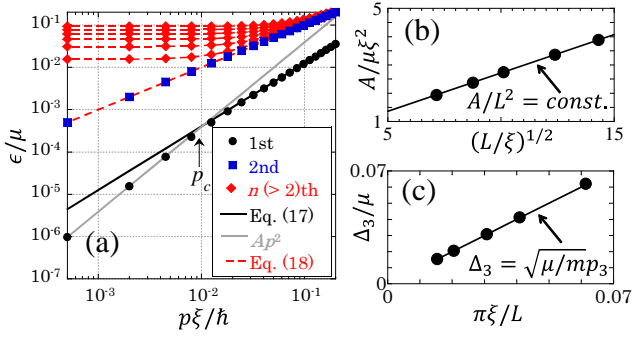}
\end{center}
\caption{(Color online) (a) The dispersion of the $n$th lowest-energy modes ($1 \leq n \leq 8$) in a finite-size system ($L=204.8\xi$).
The 1st mode has a fractional dispersion of \Eq{eq:dis_ripplon} above the critical momentum $p_c$, which satisfies $A p_c^2=\epsilon_{\rm rip}(p_c)$. The dispersion of the $n$th mode ($n \geq 3$) approaches to that of phonon (the 2nd mode) for $p \gg p_n$ but has an energy gap $\Delta_n$ for $p\to 0$. (b), (c) The system size dependence of $A$ (b) and $\Delta_3$ (c). 
} 
\label{dispersion123}
\end{figure}

To identify all NG modes in this system, we numerically diagonalized Eq.~(\ref{eq:reducedBdG})
 for low-energy modes with $p\xi/\hbar \ll 1$, paying particular attention to the finiteness of the system size.
Numerical simulations of Eq.~(\ref{eq:reducedBdG}) were done by imposing the boundary conditions,
$du_j/dz=dv_j/dz=0$ at the edges $z=\pm L/2$ of the system, and $\psi_j=\sqrt{\mu_j/g_{jj}}$ at $z=(-1)^jL/2$ and $\psi_j=0$ at $z=-(-1)^jL/2$.
Figure \ref{dispersion123} shows dispersions of the several low-energy modes for the same parameters as in \Fig{domainprofile}(a) with a system size $L=204.8 \xi$.

The first lowest mode for $p>p_c$ is well fitted to the analytic dispersion of Eq.~(\ref{eq:dis_ripplon}).
Here, we used the approximated form of the tension coefficient $\sigma=\sqrt{2}P\xi\sqrt{g_{12}/g-1}$ \cite{Schaeybroeck} for the fitting.
However, the dispersion obeys $\epsilon =A p^2$ for $p$ smaller than a certain critical value $p_c$ that depends on $L$.
We found that the coefficient $A$ is proportional to $\sqrt{L}$ [see Fig.~\ref{dispersion123}(b)], 
and thus the critical momentum $p_c$, 
at which the $p^2$ and $p^{3/2}$ branches are connected, depends on the system size $L$ as $p_c \propto 1/L$.
Therefore, the dispersion $\epsilon =A p^2$ comes from the finite-size effect, and the first lowest mode in the infinite system with $L=\infty$ coincides with the ripplon described by the effective theory.
Actually, the spatial profiles of the density and phase fluctuations on the $p^{3/2}$ branch 
are well explained with the analytic forms of Eqs.~(\ref{eq:d_n}) and (\ref{eq:d_theta}), as shown in \Fig{domainprofile}(b).

The second lowest mode is well fitted to $\epsilon =\sqrt{\mu/m} ~ p$ in the whole momentum regime, being independent of $L$.
This mode is a phonon propagating along a topological defect,
 which is an analog of a varicose mode propagating along a vortex \cite{Takeuchi2009}. 
A finite-size effect clearly appears for the $n$th lowest mode with $n \geq 3$.
This effect makes the $n$th mode with $n\geq 3$ ``massive'' with an energy gap $\Delta_n$ at $p=0$ 
and the dispersion of the $n$th mode is asymptotic to that of the second mode with increasing $p$.
This behavior is explained by the semi-classical theory with \Eq{eq:dS} as
\bea
\epsilon =\epsilon_n\equiv \sqrt{\mu/m}\sqrt{p^2+p_n^2}~~~~(n>1),
\label{phonon_n}
\eea 
where $p_n/\hbar=(n-2)\pi/L$ is the wave number of the $n$th mode normal to the wall.
The wave numbers $p_n/\hbar$ correspond to those of standing waves emerging in the phase and density fluctuations [see \Fig{domainprofile}(b)].
The standing wave is constructed with a superposition of the plane waves described by \Eq{eq:semi-U} with $S=\pm p_n z$,
 which propagate through the wall without disturbance.
 This transmission property of the Bogoliubov modes is an analog of so-called anomalous tunneling through a ferromagnetic domain wall in the low-energy limit \cite{Watabe2012}.
Equation (\ref{phonon_n}) explains very well the dispersion and the energy gap with $\Delta_n = \sqrt{\mu/m}~p_n$ in \Fig{dispersion123}(a) and (c).

\begin{figure}
\begin{center}
\includegraphics[width=0.8 \linewidth,keepaspectratio]{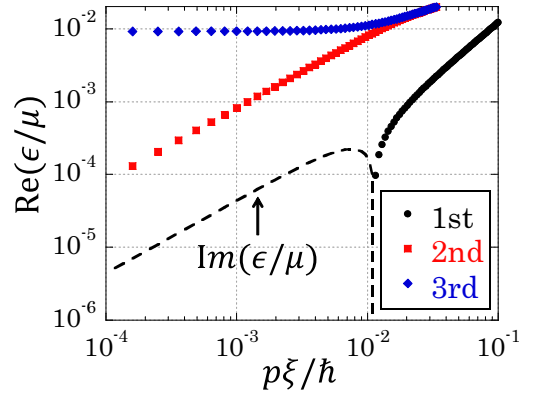}
\end{center}
\caption{(Color online) Double logarithmic plot of the dispersion of the lowest energy modes in a harmonically trapped system. The dashed curve shows the imaginary part of the dispersion of the first mode.} 
\label{fig:trap}
\end{figure}

The two NG modes are realized in experiments of atomic BECs.
To achieve the low-energy dispersions,
 the domain wall width $\xi_w$ must be much smaller than both the system sizes $L_{||}$ along a domain wall and $L_\bot$ normal to the wall.
For segregated condensates in a harmonic potential $V_j=(m/2)[\omega_{||}(x^2+y^2)+\omega_\bot^2z^2]$,
 $L_{||, \bot}$ may correspond to the Thomas-Fermi radii $R_{||,\bot}=\sqrt{2}\xi \mu/\hbar\omega_{||,\bot}$.
 Figure \ref{fig:trap} shows the dispersions of the three lowest modes in a trapped system with $R_\bot/\xi =153.6$ and $\omega_{||}\to 0$ by assuming $\pi\hbar/p \ll R_{||}$, where the parameters are the same as those used in \Fig{domainprofile}.
 Above the critical momentum $p_c$ defined by $p_c\xi/\hbar \equiv \pi\xi/R_\bot = 0.02$,
 we find the characteristic dispersion similar to \Fig{dispersion123}(a).
However, the first lowest mode has imaginary dispersion ${\rm Im}\epsilon > 0$ for $p \lesssim p_c$;
 the system may be dynamically unstable leading to amplification of ripplons.
 This is because  the total energy decreases when the wall becomes parallel to the $z$ axis for $R_{||}>R_\bot$.
Therefore, a domain wall is stabilized for $R_\bot \gtrsim R_{||}$ and then the $p$ and $p^{3/2}$ branches are realized \cite{Ticknor2013}.
The in-phase phonon (ripplon) can be excited by homogeneous (localized) perturbation potentials whose signs are in-phase (out-of-phase) between two components.
Their dispersion relations can be detected directly from the time dependencies of the emergent density fluctuations on the periodicities of the perturbation potentials along the domain wall.

 Finally, it should be mentioned that
 $n_{\rm NG}=2$ is smaller than $n_{\rm SSB}=3$ in this system.
The second lowest mode, namely the in-phase phonon, is identical to the NG mode related to the breaking of a subset of the original U(1)$\times$U(1) symmetry:
 the zero-energy perturbation with $\Delta\Theta_j=\Delta\Theta$ in \Eq{eq:zero_phase}.
Correspondingly,
 the first lowest mode, the ripplon, may be understood as a result of a pairing between the out-of-phase rotation with $\Delta\Theta_j=(-1)^j\Delta\Theta$ in \Eq{eq:zero_phase} and the domain wall shift of \Eq{eq:zero_shift};
 the density and phase perturbations caused by the ripplon in the $p\to 0$ limit mimic those by the wall shift and the out-of-phase rotation [see \Fig{domainprofile}(a) and (b) 1st].
According to Ref. \cite{Watanabe2}, the pairing of zero modes links directly to the linear independence between the modes;
 we found that the two zero-energy perturbations, namely, the out-of-phase rotation and the wall shift, are not orthogonalized to each other as
 ${\cal N}_{nn'}\neq 0$, while ${\cal N}_{nn'}= 0$ for other combinations among the three zero-energy perturbations.
We may expect a similar situation on the ferromagnetic domain wall in spinor BECs.
 To study such anomalous behaviors of zero modes for a variety of topological solitons realized in multi-component superfluid systems is an interesting direction in which to develop a universal theory of NG modes.

\begin{acknowledgments}
This work was supported by the ``Topological Quantum Phenomena'' (No. 22103003) Grant-in Aid for Scientific Research on Innovative Areas from the Ministry of Education, Culture, Sports, Science and Technology (MEXT) of Japan.
\end{acknowledgments}



\begin{thebibliography}{99}
\bibitem{Res2} Another possible solution $(dS/dz)^2=-(2\hbar/\xi)^2 $ in the limit $p\to 0$ is excluded because our theory neglects the internal structure of the wall within $|z| \lesssim \xi_w$.
\bibitem{Res1} 
In Ref. \cite{Mazets}, the $p^{1/2}$-dispersion was considered to be analogous to that of surface gravity waves in hydrodynamics.
However, in Ref. \cite{Takeuchi}, it has been theoretically formulated and numerically confirmed that the role of gravity is played by the external forces due to the potential gradient normal to the interface in segregated condensates. We have never observed the $p^{1/2}$-dispersion without external potentials in our careful numerical simulations.
\end{thebibliography}
\end{document}